\documentclass[conference]{IEEEtran}
\IEEEoverridecommandlockouts
\usepackage{optidef}
\usepackage{authblk}
\usepackage{cite}
\usepackage{booktabs}
\usepackage{tabu}
\usepackage{amsmath,amssymb,amsfonts}
\usepackage[hang,flushmargin]{footmisc}
\usepackage{multirow}
\usepackage{float}
\usepackage{textcomp}
\usepackage{lscape}
\usepackage{url}
\usepackage{enumerate}
\usepackage{caption}
\usepackage{subcaption}
\usepackage{algpseudocode}
\usepackage{graphicx}
\usepackage{textcomp}
\usepackage{xcolor}
\usepackage{array}
\usepackage{mathtools,algorithm}

\usepackage{enumitem}

\newcommand{\CASE}[1]{\STATE \textbf{case} #1\textbf{:} \begin{ALC@g}}
\newcommand{\ENDCASE}{\end{ALC@g}}

\newcommand{\DEFAULT}{\STATE \textbf{default:} \begin{ALC@g}}
\newcommand{\ENDDEFAULT}{\end{ALC@g}}
\newcommand{\DEFAULTLINE}[1]{\STATE \textbf{default:} }
\newcolumntype{L}[1]{>{\raggedright\let\newline\\\arraybackslash\hspace{0pt}}m{#1}}
\newcolumntype{C}[1]{>{\centering\let\newline\\\arraybackslash\hspace{0pt}}m{#1}}
\newcolumntype{R}[1]{>{\raggedleft\let\newline\\\arraybackslash\hspace{0pt}}m{#1}}
\def\BibTeX{{\rm B\kern-.05em{\sc i\kern-.025em b}\kern-.08em
    T\kern-.1667em\lower.7ex\hbox{E}\kern-.125emX}}

\begin{document}

\title{Application of S-band for Protection in Multi-band Flexible-Grid Optical Networks}

 \author[ ]{Varsha Lohani}
 \author[ ]{Anjali Sharma}
 \author[ ]{Yatindra Nath Singh}

 \affil[  ]{Department of Electrical Engineering, Indian Institute of Technology Kanpur, Kanpur, India}

 \affil[  ]{\textit {lohani.varsha7@gmail.com*, anjalienix05@gmail.com, ynsingh@iitk.ac.in}}

\maketitle
\begin{abstract}

The core network is experiencing bandwidth capacity constraints as internet traffic grows. As a result, the notion of a Multi-band flexible-grid optical network was established to increase the lifespan of an optical core network. In this paper, we use the C+L band for working traffic transmission and the S-band for protection against failure. Furthermore, we compare the proposed method with the existing ones.
\end{abstract}

\section{Introduction}

According to the Cisco Annual Internet Survey, nearly two-thirds of the world's population will have Internet connectivity by 2023. The number of nodes connected to the IP networks would be more than three times the number of people worldwide. In 2023, the average of fixed global broadband speeds will hit 110.4 Mbps, up from 45.9 Mbps in 2018\cite{cisco}. The evolution of high-speed technology requires a network that can satisfy bandwidth requirements. Two networks built upon optical fibre technologies have been proposed and studied in the past few decades. These are the Fixed-grid optical networks and Flexible-grid optical networks \cite{itut,eon}. 

Optical networks need to carry enormous amounts of traffic while maintaining service continuity even in faults. Failure of even a single link\footnote{span, link and edges are used interchangeably} will result in the loss of a substantial amount of data if not protected automatically. Therefore, survivability against link or path failures is an essential design requirement for high-speed optical networks. A survivability scheme aims to offer reliable services for the large volume of traffic even in the presence of failures\footnote{Fiber cut, human-made errors or natural disasters such as earthquakes, hurricanes, etc.} as well as abnormal operating conditions.

Currently, we are working with the C-band of the optical spectrum. Due to bandwidth capacity crunch at higher traffic loads, many incoming lightpath requests get blocked. To increase the bandwidth capacity of the fibre, another dimension of interest is Spatial. Spatial resources can be allocated flexibly to incoming mixed-rate traffic demands to scale up the network capacity. An effective and innovative spatial approach uses another band in the electromagnetic spectrum. The use of bands S and L along with C increases the network capacity and reduces the blocking probability of the new incoming connection requests. However, due to the S and L bands, the Generalized Optical Signal to Noise Ratio (GOSNR) is degraded in the C-band. A model of GOSNR is formulated in \cite{mb3}.

In this paper, for Routing, Spectrum Assignment \cite{RSA1,mb2,self1} and Protection Provisioning \cite{PR} in Multi-band Flexible-grid Optical Networks we consider C\footnote{1530 nm - 1565 nm}+L\footnote{1565 nm - 1625 nm} bands to carry primary traffic due to their low noise figure \cite{mb1} whereas S\footnote{1460 nm - 1530 nm} band is used for protection against failures.

\section{Proposed Scheme}
The Quality of Transmission (QoT) for lightpath provisioning is an essential parameter for a flexible-grid optical network. In this paper, QoT is modelled in terms of GOSNR as discussed in \cite{mb4}. The QoT can be estimated using GNPy, an open-source optical route planning library \cite{gnpy}.

\begin{figure*}[htbp]
    \centering
    \includegraphics[width=0.8\linewidth]{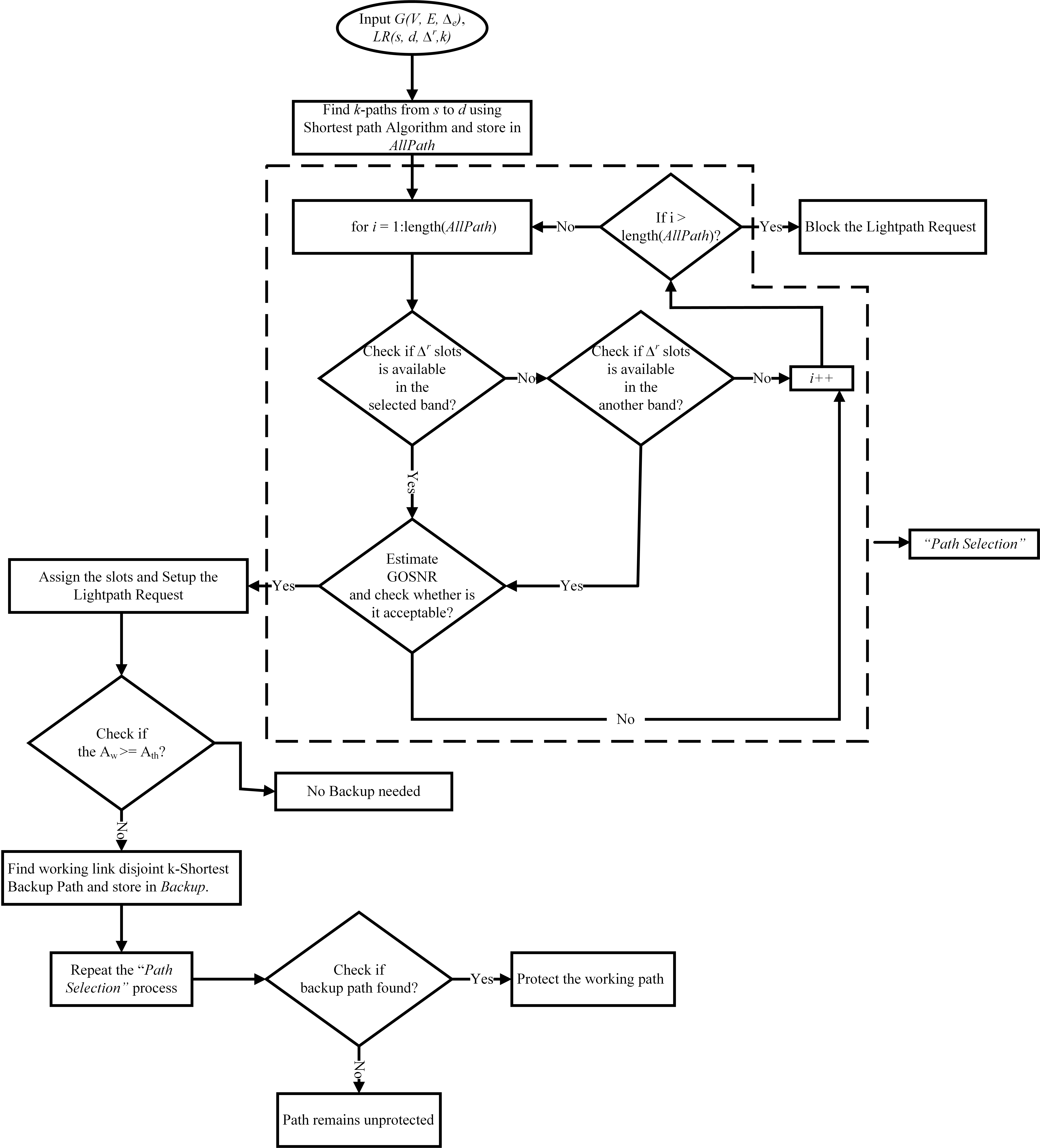}
    \caption{Flowchart of the proposed scheme.}
    \label{fig:cls}
\end{figure*}

The flowchart of the proposed method for MB flexible-grid optical networks is shown in figure \ref{fig:cls}. The terminologies used in the flowchart are defined as:
\begin{itemize}
    \item $G(V,E, \{ \Delta_{e} \})$: We represent an optical network as a graph \textit{G(V, E)} where \textit{G} is defined as a set of optical vertices \textit{V}, and set of optical fiber edges \textit{E}. Each edge $\textit{e}\in \textit{E}$ has usable slots $\Delta_{e}$.

    \item $LR(s,d,| \Delta^{r} |, k)$ is a Lightpath Request where $s$ is the source node, $s \in V$, and $d$ is the destination node, $d \in V$. In this paper, we use a grid size of 12.5 GHz and BPSK modulation format, i.e., the modulation level $m = 1$. We are also considering an additional guard band (GB) so that no two lightpaths interfere if placed next to each other. The maximum number of paths to be computed by RSA is represented by \textit{ k}, a positive integer. 
    \item $A_{w}$ is the Availability of the working lightpath. Considering each links to be independent, therefore, $A_{w} = \prod_{i=1}^{e} A_{i}$, where \textit{e} are the edges which are part of the working path \textit{w}.
    \item $A_{th}$ is the Threshold Availability. 
    \end{itemize}

Whenever a lightpath request arrives, \textit{k}-shortest path is computed using Dijkstra's Algorithm. Next, for spectrum allocation, we will check the spectrum slots availability first in the C-band and if the required number of slots are not available in the C-band, then select the L-band. Afterwards, compute the GOSNR of the lightpath request. If GOSNR is acceptable, allocate the spectrum slots and set up the lightpath request. 

Find the working link-disjoint backup path once the working path is set up. There is no need to allocate a backup path to every working path. The backup path is only set up when $A_w < A_{th}$ \cite{self2}. The process for finding a backup path is also the same, except for the backup path we consider only S-band. Here, we are using the Shared Backup Path and Slots Protection Technique. Therefore, whenever a failure happens, switching from working to backup happens at the end nodes of the lightpath.

\section{Discussion and Comparison}
In this work, according to the grid size of 12.5 GHz \cite{itut} for the working path, we have 868 spectrum slots (L-band has 548 slots and C-band has 320 slots), whereas, for the backup path, there are 732 slots (S-band). Here, we provide a comparison of the proposed method with the existing works.

In Standard Single Mode Fibre (SSMF) based Flexible-grid Optical Networks, the main problem is the bandwidth capacity crunch. To maximize the accommodation of the connection requests, the Core-based Flexible-grid Optical Networks \cite{self3} and Multi-band (MB) based Flexible-grid Optical Networks are used. The main problem with core-based flexible-grid optical networks is that as the number of cores increases, the problem of inter-symbol interference has a greater impact on the quality of transmission. Therefore, in this paper, we are considering MB-based Flexible-grid Optical Networks. 

Also, in the case of Multi-core Optical for Routing and Spectrum Assignment, we need to follow spectrum contiguity, continuity, and core continuity constraints. This additional constraint increases switch complexities within the network. Additionally, the protection in core-based networks can provide independent cores than Routing and Spectrum Assignment. However, core continuity needs to be followed within different cores used for protection. Whereas there is no additional constraint in Multi-band, only spectrum contiguity and continuity constraints need to be followed. 

Core-based multiband optical networks can be designed to further improve the capacity of multiband optical networks. However, merging two spatial components can increase the complexity of designing the optical nodes. 

\section{Conclusion}
In this paper, we present a scheme for Routing, Spectrum Assignment and Protection Provisioning in Multi-band Flexible-grid Optical Networks. Here, we consider C+L bands to carry direct traffic due to their low noise figure, whereas S-band is used for survivability against failures. Furthermore, we also compare the proposed scheme with conventional single-mode fibre and core-based fibre schemes.

\end{document}